\documentclass[prd,aps,showpacs,tightline,twocolumn,nofootinbib,superscriptaddress]{revtex4}
\usepackage{graphicx}
\usepackage{amssymb}
\usepackage{amsmath}

\newcommand{\bear}{\begin{array}}  
\newcommand {\eear}{\end{array}}
\newcommand{\bea}{\begin{eqnarray}}   
\newcommand{\eea}{\end{eqnarray}}
\newcommand{\beq}{\begin{equation}}   
\newcommand{\eeq}{\end{equation}}
\newcommand{\bef}{\begin{figure}}  \newcommand 
{\eef}{\end{figure}}
\newcommand{\bec}{\begin{center}}  \newcommand 
{\eec}{\end{center}}

\begin{document}



\preprint{ICRR-Report-545}
\preprint{IPMU 09-0063}

\title{
Upward muon signals at neutrino detectors as a probe of dark matter properties
}

\author{Junji Hisano}
\affiliation{Institute for Cosmic Ray Research, University of Tokyo, Kashiwa 277-8582, Japan}
\affiliation{Institute for the Physics and Mathematics of the Universe, University of Tokyo, Kashiwa 277-8568, Japan}

\author{Kazunori Nakayama}
\affiliation{Institute for Cosmic Ray Research, University of Tokyo, Kashiwa 277-8582, Japan}

\author{Masaki J.S. Yang}
\affiliation{Institute for Cosmic Ray Research, University of Tokyo, Kashiwa 277-8582, Japan}

\date{\today}

\begin{abstract} 
We study upward muon flux at neutrino detectors such as Super-Kamiokande
resulting from high-energy neutrinos produced by the dark matter annihilation/decay
at the Galactic center.
In particular, we distinguish showering and non-showering muons
as their energy loss processes inside the detector,
and show that this information is useful for discriminating dark matter models.
\end{abstract}


\maketitle


\section{Introduction}

Recent observations of cosmic ray positron/electron fluxes 
by the PAMELA satellite~\cite{Adriani:2008zr} and ATIC balloon experiments~\cite{:2008zz}
have shown excess of events compared with expected astrophysical backgrounds.
PPB-BETS~\cite{Torii:2008xu}, which is another balloon experiment,
also reported cosmic ray electron flux consistent with ATIC results.
Although more recent results of Fermi satellite~\cite{Collaboration:2009zk} and 
HESS telescope~\cite{Aharonian:2009ah}
on the electron flux did not confirm such a sharp excess observed by ATIC,
the positron excess observed by PAMELA still remains viable.
While these anomalous positron/electron fluxes may be explained by some
nearby high-energy astrophysical sources~\cite{Atoian:1995ux,Hooper:2008kg},
they can also be interpreted as evidence of dark matter annihilation/decay~\cite{Bergstrom:2008gr}.
If the recent Fermi data is fitted by dark matter annihilation/decay, a heavy dark matter
with mass of order a few TeV is favored~\cite{Meade:2009iu}.
In the case of annihilation, the required annihilation cross section for reproducing 
the PAMELA anomaly is around $\langle \sigma v\rangle \sim 10^{-23}~{\rm cm^3s^{-1}}$,
which is orders of magnitude larger than the standard value, $3\times 10^{-26}~{\rm cm^3s^{-1}}$.
In order to fill this gap, one needs to invoke nonthermal dark matter production scenario
\cite{Kawasaki:1995cy,Moroi:1999zb,Profumo:2004ty} 
or Sommerfeld enhancement mechanism~\cite{Hisano:2003ec}
giving velocity-dependent annihilation rate.
In the case of decay, the required decay rate is about $\Gamma \sim 10^{-26}~{\rm s}^{-1}$.

In order to confirm that the positron/electron excesses are caused by dark matter annihilation/decay,
cross checks by other observations are necessary.
Gamma-rays~\cite{Bertone:2008xr}, 
anti-protons~\cite{Donato:2008jk}, 
anti-deuterons~\cite{Braeuninger:2009pe}
and radio emission associated with
the same annihilation/decay processes~\cite{Cumberbatch:2009ji}
may be used for this purpose.
At the present stage, however, no clear excess was found on those observations,
and they give constraints on annihilating/decaying dark matter models.
Big-bang nucleosynthesis \cite{Jedamzik:2004ip,Hisano:2008ti} and the reionization~\cite{Galli:2009zc} also give stringent constraints on annihilating dark matter models.

It was also pointed out that dark matter annihilation/decay which explains the PAMELA anomaly
may predict potentially large neutrino flux as well~\cite{Hisano:2008ah,Liu:2008ci},
which are constrained from observations of Super-Kamiokande (SK)
toward the direction of the Galactic center~\cite{Desai:2004pq,Desai:2007ra}.
Dark matter annihilation (decay) yields high-energy neutrinos at the Galactic center
where the number density is enhanced,
and they reach to the Earth without energy loss.
During the propagation, the neutrino oscillation effects mix three flavors ($\nu_e,\nu_\mu,\nu_\tau$).
Among them, some fraction of muon neutrinos are converted into high-energy muons 
due to the interaction with matter inside the Earth, and
if the interaction point is within around 1~km from the detector,
produced muons penetrate the detector leaving characteristic signatures
like the Cherenkov light from a muon itself 
and/or from electromagnetic shower produced by it~\cite{Ritz:1987mh,Kamionkowski:1991nj}.
Since the direction of a primary neutrino is conserved even after it is converted into a muon,
SK limit on the muon flux from the Galactic center direction provides
useful constraint on dark matter 
annihilation/decay models~\cite{Hisano:2008ah,Mardon:2009rc,Meade:2009iu}.

In this paper we further study the neutrino flux from the Galactic center generated by
dark matter annihilation/decay and how they are detected at the neutrino detector.
Since both conversion rate of neutrino into muon 
and muon energy loss processes in the Earth matter depend on its energy,
the signal of the high-energy neutrino has energy dependence.
In particular, we distinguish {\it showering} and {\it non-showering} muon events at SK,
following the criterion given in Ref.~\cite{Desai:2007ra}.
As is expected, higher energy muons are likely to lose energy by radiative losses,
developing electromagnetic shower inside the detector.
Interestingly, the probability for causing shower events becomes significant
around the muon energy $\sim 1~{\rm TeV}$, which is a natural scale of dark matter mass. 
Therefore such a separation, showering/non-showering events, has rich information
on the initial neutrino energy spectrum and
we show that it is a possible method to discriminate dark matter properties.\footnote{
The importance of a separation of showering and non-showering events was also pointed out in
Ref.~\cite{Meade:2009iu}.
}
This analysis can be applied to future neutrino detectors, such as KM3NeT, for
searching for dark matter, or even the present on-going experiments, such as SK,
may be able to improve constraints on dark matter properties.

In Sec.~\ref{sec:muon} we provide formulae to evaluate the upward 
showering and non-showering muon fluxes from the primary neutrino flux generated by 
dark matter annihilation/decay at the Galactic center.
Applications to some dark matter annihilation/decay models
is performed in Sec.~\ref{sec:DMmodel}. 
Sec.~\ref{sec:conc} is devoted to conclusions.

\section{Muon flux from annihilating/decaying dark matter} \label{sec:muon}

In this section we develop a formalism to evaluate the upward muon flux
starting from the initial neutrino energy spectrum produced by dark matter annihilation/decay.

\subsection{Primary neutrino flux}

We assume that DM particle, denoted by $\chi$, with mass $m$
annihilates (decays) into some
final state $F$ with annihilation cross section (decay rate)
$\langle \sigma v\rangle _F$ ($\Gamma_F$).
The primary neutrino flux from the Galactic center is given by
\begin{equation}
	\frac{dF_{\nu_\mu}}{dE_{\nu_\mu}}=\frac{R_\odot \rho_\odot^2}{8\pi m^2}
		\left (\sum_F \langle \sigma v\rangle _F \frac{dN_F^{(\nu_\mu)}}{dE_{\nu_\mu}} \right )
		\langle J_2 \rangle_\Omega \Delta \Omega,
\end{equation}
for the case of annihilation, and
\begin{equation}
	\frac{dF_{\nu_\mu}}{dE_{\nu_\mu}}=\frac{R_\odot \rho_\odot}{4\pi m}
		\left (\sum_F \Gamma _F \frac{dN_F^{(\nu_\mu)}}{dE_{\nu_\mu}} \right )
		\langle J_1 \rangle_\Omega \Delta \Omega,
\end{equation}
for the case of decay, with
\begin{equation}
	\frac{dN_F^{(\nu_\mu)}}{dE_{\nu_\mu}} = \sum_i 
	\left(P_{\nu_i \nu_\mu} \frac{dN_F^{(\nu_i)}}{dE} \right )_{E=E_{\nu_\mu}},
\end{equation}
where $i=e,\mu,\tau$,
and $P_{\nu_i\nu_\mu}$ denotes the probability that the
$\nu_i$ at the production is observed as $\nu_\mu$ at the Earth due to
the effect of neutrino oscillation and given as
$P_{\nu_e\nu_\mu}=0.22, P_{\nu_\mu \nu_\mu}=0.39$, and $P_{\nu_\tau \nu_\mu}=0.39.$
Here, $R_\odot = 8.5$~kpc and
$\rho_\odot=0.3~{\rm GeV cm^{-3}}$ are the distance of the solar
system from the Galactic center and the local dark matter density near the
solar system, respectively. $F$ collectively denotes the primary annihilation/decay modes
(e.g., $\mu^+ \mu^-$, etc.), and $dN_F^{(\nu_i)}/dE$ represents the
$\nu_i$ spectrum arising from the final state $F$
which is simulated by PYTHIA package~\cite{Sjostrand:2006za}.
The dependence on
the dark matter halo density profile is contained in the remaining
factor $\langle J_n \rangle_\Omega$, defined by
\begin{equation}
	\langle J_n \rangle_\Omega = \int \frac{d\Omega}{\Delta \Omega}
	\int_{\rm l.o.s.}\frac{dl(\psi)}{R_\odot}\left ( \frac{\rho(l)}{\rho_\odot} \right )^n,
\end{equation}
where $l(\psi)$ is the distance from us along the direction $\psi$,
which is the cone-half angle from the Galactic center within the range $0<\psi<\psi_{\rm max}$, and 
$\Delta \Omega(\equiv 2\pi (1-\cos\psi_{\rm max}))$ 
is the solid angle over which the neutrino flux is averaged.
Dark matter halo profile is characterized as follows:
\begin{equation}
	\rho(r) = \frac{\rho_0}{(r/R)^\gamma[ 1+(r/R)^\alpha ]^{(\beta-\gamma)/\alpha}},
\end{equation}
where $r$ is the distance from Galactic center.
Here we consider two typical density profiles: Navarro-Frenk-White (NFW) profile~\cite{Navarro:1995iw}
($\alpha=1,\beta=3,\gamma=1,R=20~$kpc)
and cored isothermal profile
($\alpha=2,\beta=2,\gamma=0,R=2.8~$kpc).
Typical values of $\langle J_n \rangle_\Omega \Delta \Omega$
for these profiles are summarized in Table~\ref{table:J}.
As is expected, dark matter halo with large core~\cite{Burkert:1995yz}
predicts smaller neutrino flux, but
the dependence on the halo profile is not so large compared with 
gamma-ray flux from the Galactic center~\cite{Bertone:2008xr},
as long as the angular resolution is larger than $\sim 5^\circ$.
Thus following discussion does not much depend on the halo profile
and it is possible that neutrino signals are detectable while the
gamma-ray constraints are satisfied.

\begin{table}[t]
  \begin{center}
    \begin{tabular}{ l | l l l l l}
      \hline \hline
      $\langle J_2 \rangle_\Omega \Delta \Omega$ 
      & $5^\circ$
      & $10^\circ$
      & $15^\circ$
      & $20^\circ$
      & $25^\circ$
      \\
      \hline
      NFW& 6.0 & 10  & 14 & 17 & 20 \\
      isothermal~~~& 1.3  & 4.3 & 8.0  & 11 & 15 \\
      \hline \hline
      $\langle J_1 \rangle_\Omega \Delta \Omega$ 
      & $5^\circ$
      & $10^\circ$
      & $15^\circ$
      & $20^\circ$
      & $25^\circ$
      \\
      \hline
      NFW& 0.3 & 1.0  & 1.9 & 2.9 & 4.0 \\
      isothermal~~~& 0.2  & 0.8 & 1.7  & 2.8 & 4.0 \\
      \hline
    \end{tabular}
    \caption{Typical values of $J$-factor for NFW and isothermal profiles for
     $\psi_{\rm max}=5^\circ,10^\circ,15^\circ,20^\circ$ and $25^\circ$. }
    \label{table:J}
  \end{center}
\end{table}

\subsection{Upward muon flux}

High-energy neutrinos produced by the dark matter annihilation/decay at the Galactic center
come in the Earth and some of them interact with matter inside the Earth,
generating high-energy muons.
Since muons have electric charges, they lose their energy during the propagation inside the Earth.
Typical propagation distance of high-energy muons with energy $\sim 1$~TeV is around 1~km,
as will be shown.
Therefore, muons appearing only within 1~km around the detector can reach the detector,
leaving their characteristic signals.
The search strategy is to find upward muons in order to avoid cosmic-ray muon background. 

Taking these processes into account, 
showering muon flux from the Galactic center within cone-half angle $\theta$ is given by
\begin{equation}
	N_{\mu^+\mu^- }^{\rm (shower)} = \int dE_{\nu_\mu} \frac{dF_{\nu_\mu}}{dE_{\nu_\mu}}
		f(E_{\nu_\mu}) \epsilon(E_{\nu_\mu}),  \label{shower}
\end{equation}
where $dF_{\nu_\mu}/dE_{\nu_\mu}$ is the parent differential neutrino flux
evaluated in the previous subsection,
$E_{\nu_\mu}$ is the parent neutrino energy,
$f(E_{\nu_\mu})$ denotes the probability that a neutrino with energy $E_{\nu_\mu}$
is converted into a muon with observable energy range,
and $\epsilon(E_{\nu_\mu})$ is fraction of the showering muon in the total observed muon events
with parent neutrino energy $E_{\nu_\mu}$.

\begin{figure}[t]
 \begin{center}
   \includegraphics[width=1.0\linewidth]{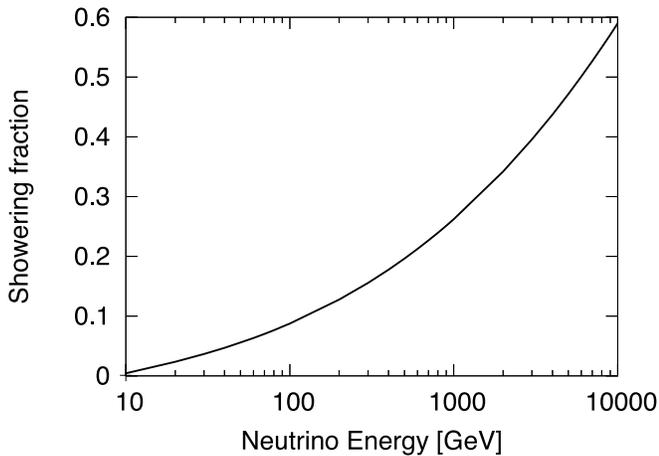} 
   \caption{ Showering fraction $\epsilon(E_{\nu_\mu})$ as a function of 
   initial neutrino energy $E_{\nu_\mu}$ given in Ref.~\cite{Desai:2007ra}.}
   \label{fig:showprob}
 \end{center}
\end{figure}

The last ingredient, $\epsilon(E_{\nu_\mu})$, is read off from Fig.~5 of Ref.~\cite{Desai:2007ra}.
By taking the ratio between black and black plus red lines in that figure,
we obtain a following fitting formula,
\begin{equation}
	\epsilon(E_{\nu_\mu}) = a+bx+cx^2+dx^3,
\end{equation}
where $x=\log_{10}(E_{\nu_\mu}/1{\rm GeV})$ and 
$a = -0.0491232, b = 0.059575,c = -0.0162841$, and $d = 0.0103345$.
(Note that this differs from the quantity $\epsilon (E_\mu)$ used in Ref.~\cite{Desai:2007ra}.)
The result is shown in Fig.~\ref{fig:showprob}.
Note that this formula is valid for $10~{\rm GeV}<E_{\nu_\mu}<10~{\rm TeV}$.
If we want to know total muon events, including non-showering events,
we have only to replace  $\epsilon(E_{\nu_\mu})$ with 1:
\begin{equation}
	N_{\mu^+\mu^- }^{\rm (total)} = \int dE_{\nu_\mu} \frac{dF_{\nu_\mu}}{dE_{\nu_\mu}}
		f(E_{\nu_\mu}).  \label{total}
\end{equation}


\begin{figure}[t]
 \begin{center}
   \includegraphics[width=1.0\linewidth]{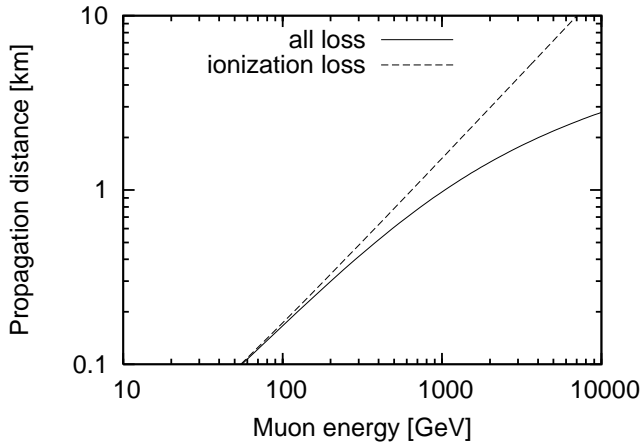} 
   \caption{Propagation distance of muon in the standard rock as a function of muon energy,
   	when all energy loss processes are taken into account (solid line)
	and only ionization loss is considered (dashed line). }
   \label{fig:R_mu}
 \end{center}
\end{figure}

In order to evaluate $f(E_{\nu_\mu})$, the muon-nucleon scattering cross section and
the muon energy loss in the rock surrounding the SK must be known.
It is given by
\begin{equation}
\begin{split}
	f(E_{\nu_\mu}) = &\int _{E_{\rm th}}^{E_{\nu_\mu}} dE_\mu
	\left( \frac{d\sigma_{\nu_\mu (\bar \nu_\mu) p\to \mu (\bar \mu)X}}{dE_\mu} 
	n^{\rm (rock)}_p \right. \\
	&+ \left. \frac{d\sigma_{\nu_\mu (\bar \nu_\mu) n\to \mu (\bar \mu)X}}{dE_\mu} 
	n^{\rm (rock)}_n \right )
	R(E_\mu,E_{\rm th}),
\end{split}
\end{equation}
where $n_{p(n)}^{\rm (rock)}$ is the number density of proton (neutron) in the rock and
$d\sigma_{\nu_\mu (\bar \nu_\mu) p (n)\to \mu X}/dE_\mu$ is the scattering cross section
of a neutrino with proton (neutron)
to create a muon with energy $E_\mu$, given by~\cite{Barger:2007xf}\footnote{
	This includes both processes such as 
	$\nu_\mu p \to \mu X$ and $\bar \nu_\mu p \to \bar \mu X$.
	Thus primary neutrino flux in Eq.~(\ref{shower}) should only count neutrinos, not anti-neutrinos
	(assuming that primary flux of neutrinos and anti-neutrinos are equal).
}
\begin{gather}
	\frac{d\sigma_{\nu_\mu (\bar \nu_\mu) p\to \mu (\bar \mu)X}}{dE_\mu}=
	\frac{2m_pG_F^2}{\pi}\left ( 0.21+ 0.29\frac{E_\mu^2}{E_{\nu_\mu}^2} \right ),\\
	\frac{d\sigma_{\nu_\mu (\bar \nu_\mu) n\to \mu (\bar \mu)X}}{dE_\mu}=
	\frac{2m_nG_F^2}{\pi}\left ( 0.29+ 0.21\frac{E_\mu^2}{E_{\nu_\mu}^2} \right ).
\end{gather}
The mass density of the standard rock is $2.65~{\rm g cm^{-3}}$ \cite{Dutta:2000hh},
and hence $n_p^{\rm (rock)}=n_n^{\rm (rock)}=2.65N_A/2~{\rm cm^{-3}}$ 
where $N_A=6.022\times 10^{23}$ is the Avogadro's number.
$R(E_\mu,E_{\rm th})$ is the distance which muons with energy $E_\mu$ can travel
until their energies decrease to the threshold energy $E_{\rm th}$ in the rock.
It is obtained by solving the energy loss equation,
\begin{equation}
	\frac{dE_\mu}{dX} = -\alpha (E_\mu) -\beta(E_\mu) E_\mu,
\end{equation}
where $X$ represents the distance in units of ${\rm g cm^{-2}}$.
The first term $\alpha (E_\mu)$ is from ionization loss, and it only logarithmically depends on $E_\mu$.
The second term denotes other radiative energy loss effects due to bremmstrahlung and pair creation,
which may cause electromagnetic shower inside the detector.
The precise formulae for those energy losses are given in Ref.~\cite{Dutta:2000hh}.
In Fig.~\ref{fig:R_mu} we show the propagation distance of muons in the rock.
For reference, we also have shown the result when only the ionization loss is taken into account.
For convenience, we provide fitting formula for $R(E_\mu,E_{\rm th}=10~{\rm GeV})$ 
in the rock, which is valid for $10~{\rm GeV}< E_\mu < 10~{\rm TeV}$, as
\begin{equation}
	R(E_\mu,E_{\rm th}=10~{\rm GeV}) = 10^{a+by+cy^2}~{\rm [km]},
\end{equation}
where $y=\log_{10}(E_\mu/1{\rm GeV})$ and 
$a = -3.29186,b = 1.52594$, and $c = -0.147224$.


Notice that both the interaction cross section of neutrinos with nucleon
and propagation distance of muons are roughly proportional to $E_{\nu_\mu}(\sim E_\mu)$.
Therefore heavier dark matter has a tendency to provide more muon signals.
This situation is quite different from other indirect signals, such as gamma-rays or positrons.
For fixed annihilation cross section or decay rate,
smaller fluxes of gamma-ray or positron will be predicted for heavier dark matter,
since the number density of dark matter is inversely proportional to the dark matter mass.
In the case of neutrino-induced muon signals, however, higher energy neutrinos 
more likely produce muon signals, so that it compensates the suppression from
smaller number density of dark matter for heavier mass~\cite{Hisano:2008ah}.

\subsection{Atmospheric background}

Before going into predictions of dark matter models,
we evaluate the background muon events coming from atmospheric neutrinos.
Since we are interested in the upward muon events, 
we do not need to worry about background from atmospheric muons.
Detailed calculations for the atmospheric neutrino spectrum are performed in 
Refs.~\cite{Honda:2004yz,Honda:2006qj}.
Using the atmospheric neutrino flux given in Ref.~\cite{Honda:2006qj},
the background upward muon events are evaluated as
$N_{\mu^+\mu^-}^{(\rm total,BG)}\simeq 2.8\times 10^{-15}~{\rm cm^{-2}s^{-1}}$ and
$N_{\mu^+\mu^-}^{(\rm shower,BG)}\simeq 0.34\times 10^{-15}~{\rm cm^{-2}s^{-1}}$
from the cone half angle $\psi_{\rm max}=5^\circ$ (corresponding to 
$N_{\mu^+\mu^-}^{(\rm total,BG)}\simeq 1.2\times 10^{-13}~{\rm cm^{-2}s^{-1}sr^{-1}}$ and
$N_{\mu^+\mu^-}^{(\rm shower,BG)}\simeq 1.4\times 10^{-14}~{\rm cm^{-2}s^{-1}sr^{-1}}$,
respectively).
If dark matter prediction is larger than or comparable to those values, 
dark matter detection by neutrino-induced muon signals is promising.

\section{Discriminating dark matter models} \label{sec:DMmodel}

Now let us evaluate the upward showering/non-showering muon flux
following the procedure given in the previous section.
We consider dark matter mainly annihilating (decaying) into 
$\mu^+\mu^-, \tau^+\tau^-,W^+W^-,ZZ$, and $\nu_i \bar \nu_i~(i=e,\mu,\tau)$.
Following results are not changed so much between $\mu^+\mu^-$ and $\tau^+\tau^-$,
and hence we call these models as ``leptonic annihilation (decay)''.
Similarly, annihilating (decaying) into $W^+W^-$ and $ZZ$ lead to almost the same results,
and we call those cases as ``hadronic annihilation (decay)''.
Taking account of neutrino oscillation effects, annihilating (decaying) into 
$\nu_\mu \bar \nu_\mu$ and $\nu_\tau \bar \nu_\tau$ yield completely the same muon flux,
but the case of $\nu_e \bar \nu_e$ predicts slightly smaller flux.
In the following we call the former case as ``neutrino annihilation (decay)'',
since the characteristic behaviors are essentially the same.

\begin{figure}
 \begin{center}
   \includegraphics[width=1.0\linewidth]{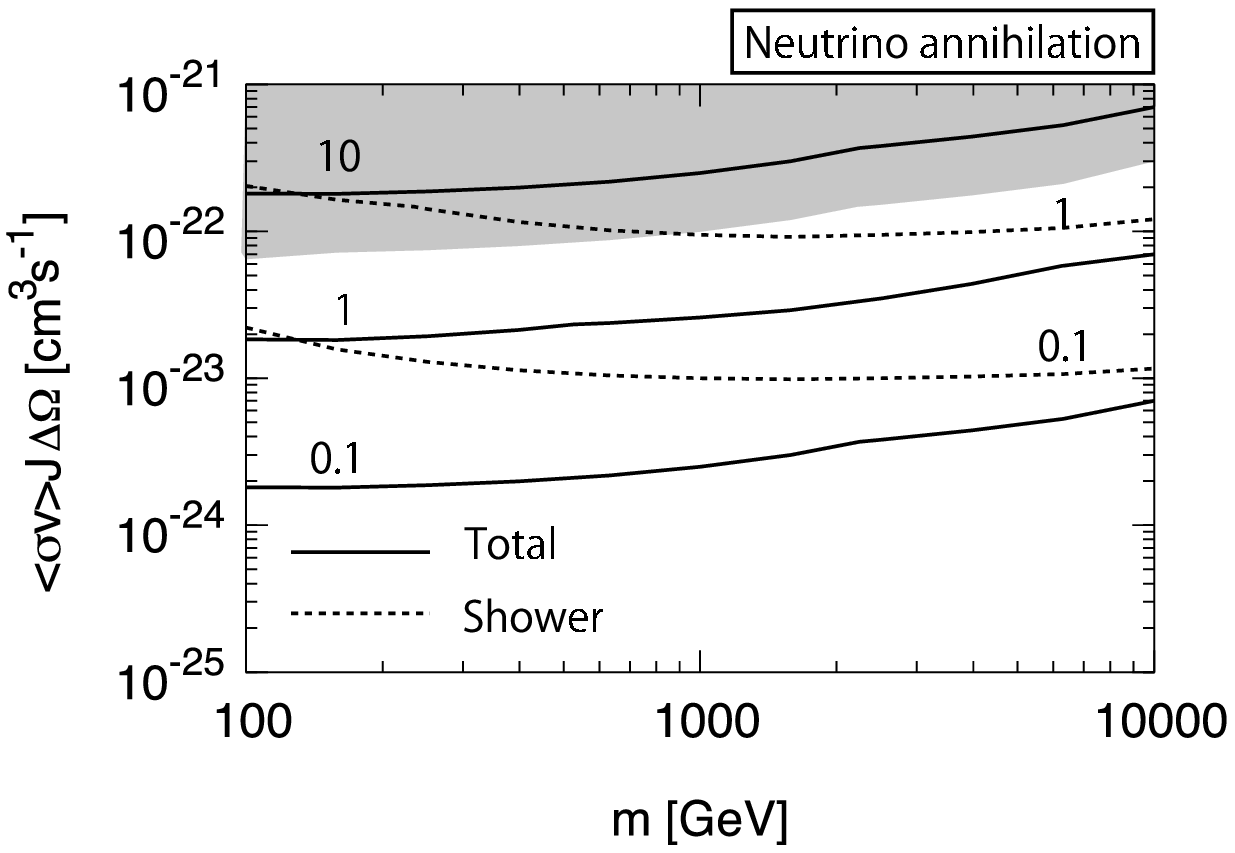} 
   \includegraphics[width=1.0\linewidth]{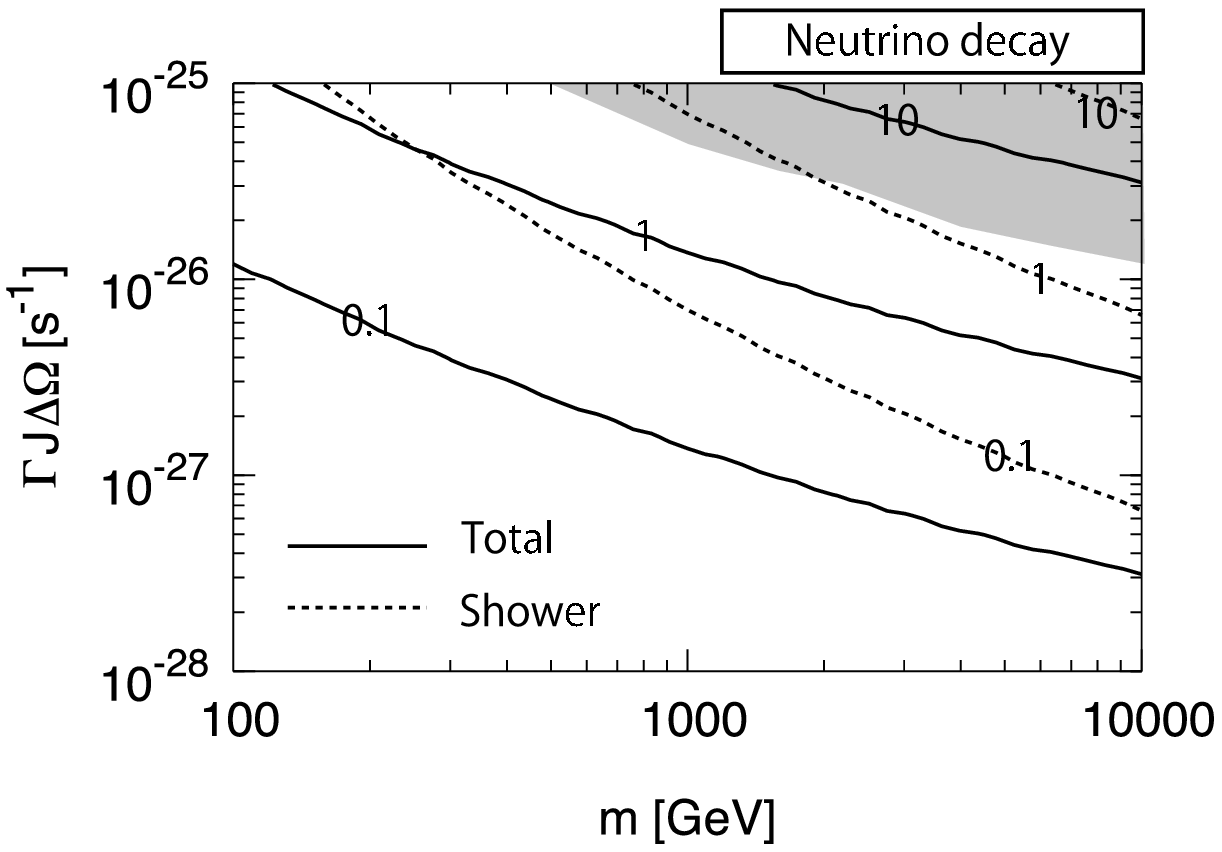} 
   \caption{ Contours of  
    $N_{\mu^+\mu^-}^{\rm (total)}=(0.1,1,10)\times 10^{-15}~{\rm cm^{-2}s^{-1}}$
    and $N_{\mu^+\mu^-}^{\rm (shower)}=(0.1,1,10)\times 10^{-15}~{\rm cm^{-2}s^{-1}}$ are shown by
    solid and dashed lines, respectively. The horizontal axis is dark matter mass and the vertical axis is
     $\langle \sigma v \rangle \langle J_2 \rangle_\Omega \Delta \Omega$
	for the case of annihilation (top), and
	$\Gamma \langle J_1 \rangle_\Omega \Delta \Omega$
	for the case of decay (bottom). 
	Here dark matter is assumed to annihilate (decay) into $\nu_\mu \bar \nu_\mu$. 
   	Results in the case of $\nu_\tau \bar \nu_\tau$ are the same.
   	In the case of  $\nu_e \bar \nu_e$, both fluxes are slightly reduced by an amount of
	factor 2. 
	The shaded region is excluded at $90\%$ C.L. by the observation of SK toward the direction of 
	Galactic center with cone half angle $5^\circ$.}
   \label{fig:nu}
 \end{center}
\end{figure}

\begin{figure}
 \begin{center}
   \includegraphics[width=1.0\linewidth]{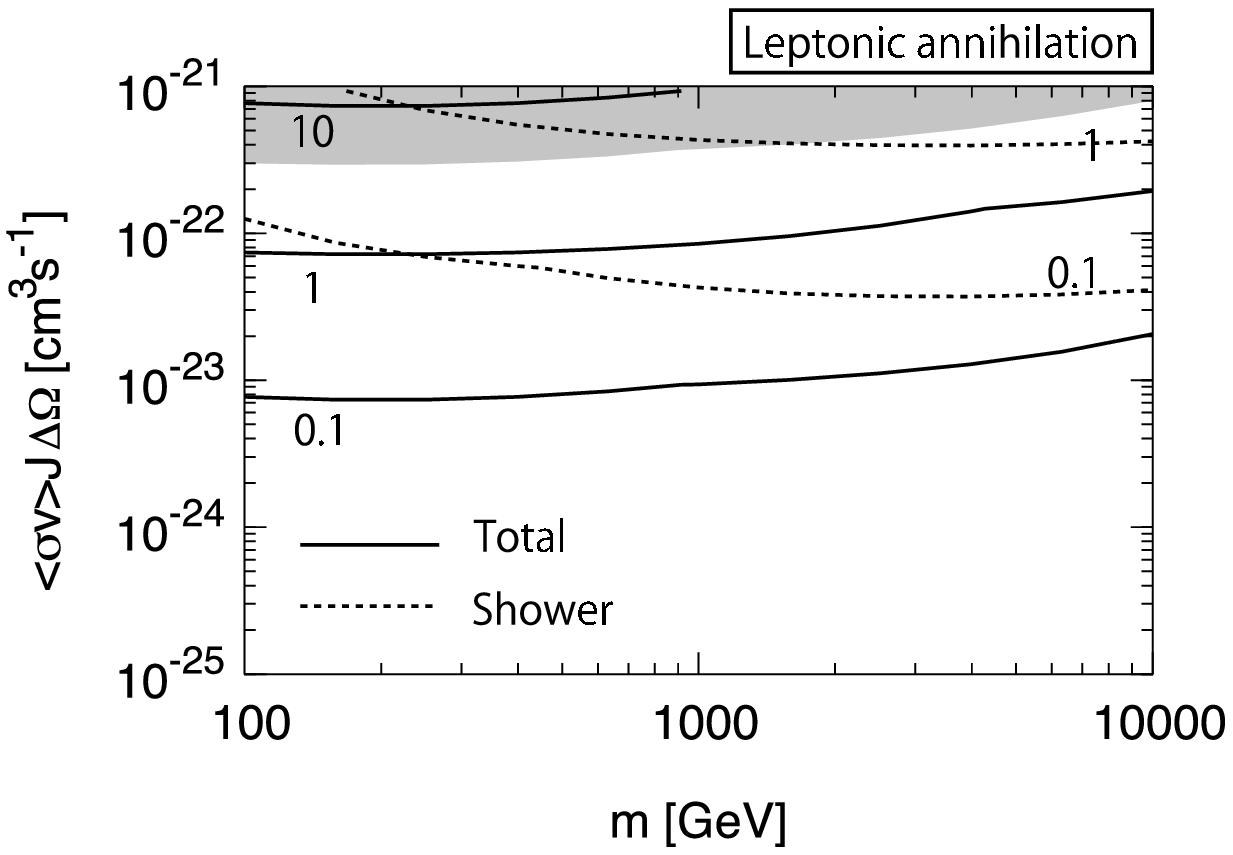} 
   \includegraphics[width=1.0\linewidth]{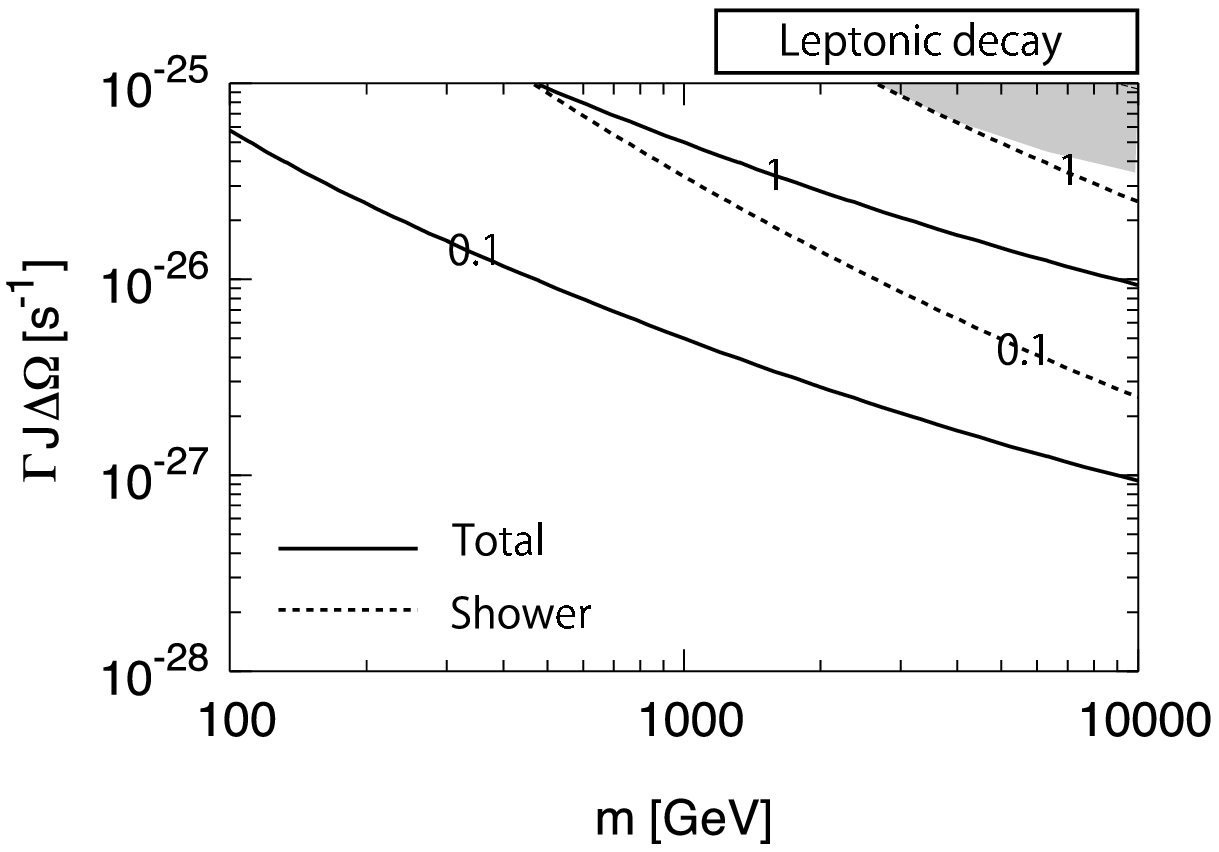} 
   \caption{ Same as Fig.~\ref{fig:nu}, but for annihilating (decaying) into $\mu^+\mu^-$. 
   	In the case of $\tau^+ \tau^-$, muon fluxes are slightly smaller by an amount of
	$\sim 10\%$. }
   \label{fig:lep}
 \end{center}
\end{figure}

\begin{figure}
 \begin{center}
   \includegraphics[width=1.0\linewidth]{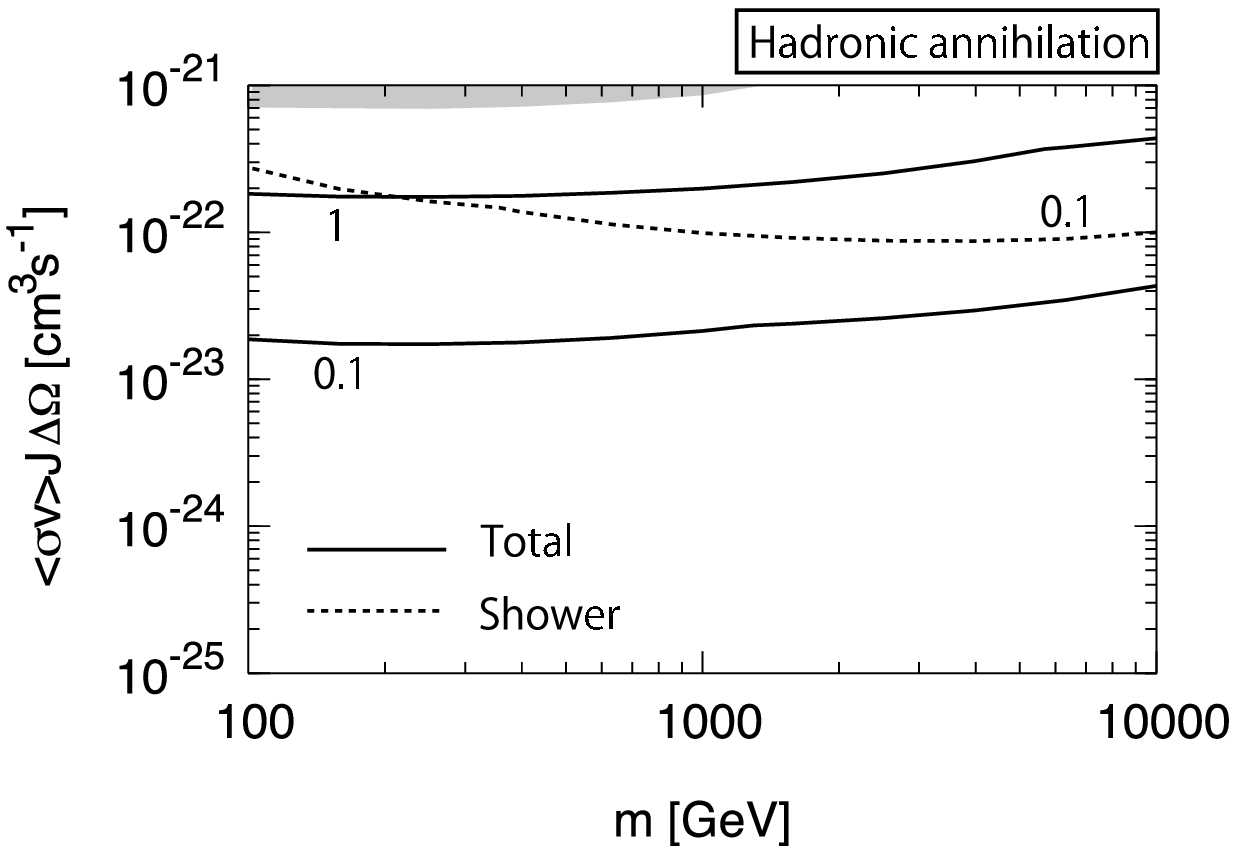} 
   \includegraphics[width=1.0\linewidth]{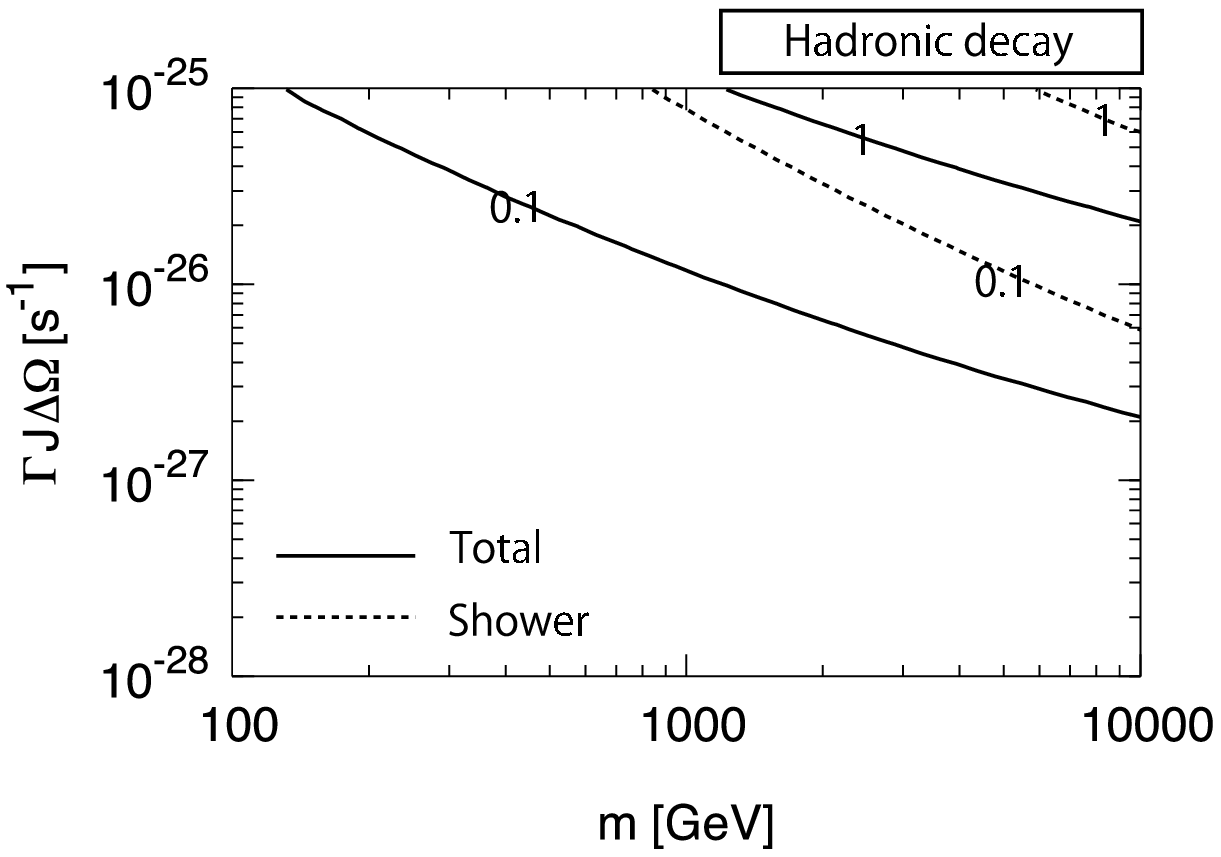} 
   \caption{ Same as Fig.~\ref{fig:nu}, but for annihilating (decaying) into $W^+W^-$. 
   	Results in the case of $ZZ$ are almost the same.}
   \label{fig:had}
 \end{center}
\end{figure}

Resulting total and showering muon fluxes are shown in Figs.~\ref{fig:nu}-\ref{fig:had},
for the above mentioned three cases: neutrino, leptonic, and hadronic annihilation (decay).
In these figures, the vertical axis is 
$\langle \sigma v \rangle \langle J_2 \rangle_\Omega \Delta \Omega$
for the case of annihilation (top panels), and
$\Gamma \langle J_1 \rangle_\Omega \Delta \Omega$
for the case of decay (bottom panels).
Contours of $N_{\mu^+\mu^-}^{\rm (total)}=(0.1,1,10)\times 10^{-15}~{\rm cm^{-2}s^{-1}}$ and
$N_{\mu^+\mu^-}^{\rm (shower)}=(0.1,1,10)\times 10^{-15}~{\rm cm^{-2}s^{-1}}$ are shown by
solid and dashed lines, respectively.
The shaded region is excluded at $90\%$ C.L. by the observation of SK toward the direction of 
Galactic center with cone half angle $5^\circ$~\cite{Desai:2004pq}.
If other values of cone angle are chosen, the constrained region becomes 
widened or narrowed.


As is easily seen from these figures, in the case of annihilation,
the total muon flux remains roughly constant
even if dark matter mass is increased for fixed annihilation cross section.
This is because the annihilation rate is proportional to $m^{-2}$ reflecting the fact that
the rate is proportional to the square of dark matter number density, but
the muon production cross section inside the Earth and the muon propagation distance are both roughly proportional to
$m$, which cancels the $m$ dependence.
However, the muon propagation distance deviates from the simple scaling $R(E_\mu)\propto E_\mu$
when the radiative energy loss becomes important at $E_\mu \gtrsim 500$~GeV 
(see Fig.~\ref{fig:R_mu}).
Thus for heavy dark matter with $m \gtrsim 500~$GeV, the predicted total muon flux
shows mass dependence.
On the other hand, the showering muon is caused by high-energy component of the primary
neutrino flux, and hence
its mass dependence is quite different from the total muon flux, although the flux itself is 
suppressed.
However, the showering muon flux prefers heavier dark matter mass,
and atmospheric background flux is significantly reduced for higher energy region.
Thus we may have a chance to detect muon signals and observe separately 
non-showering and showering muons.

Mass scaling in the case of decaying dark matter can also be recognized in a similar way.
Since the decay rate per unit volume is only proportional to the dark matter number density,
the total muon flux is proportional to $m$ for fixed decay rate.
This feature can be read off from the figures.


\begin{figure}[t]
 \begin{center}
   \includegraphics[width=1.0\linewidth]{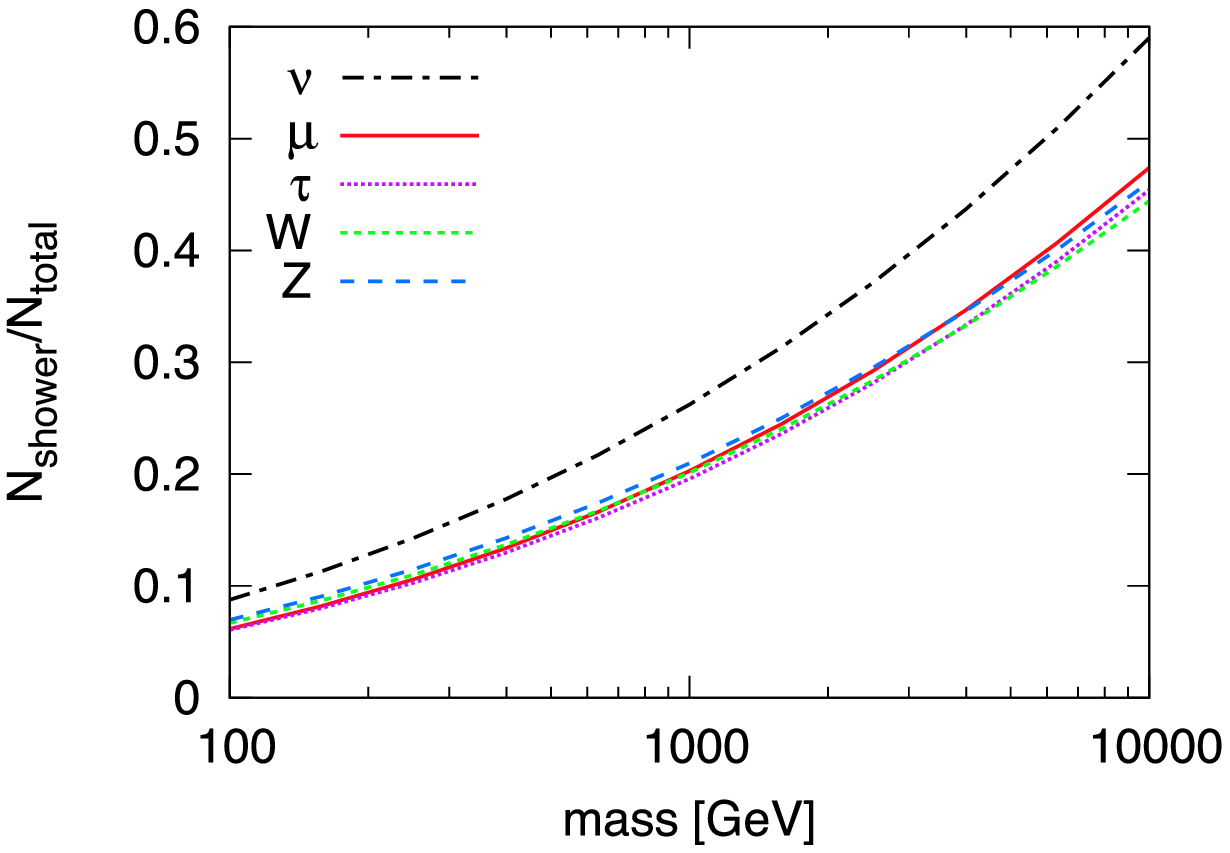} 
   \caption{Ratio of the showering muon to the total muon fluxes as a function of dark matter mass 
     in the dark matter annihilation. In the case of decay, the horizontal axis corresponds to half of the 
     dark matter mass. The label shows the
     dominant mode of the dark matter annihilation/decay: $\nu \bar
     \nu, \mu^+\mu^-,\tau^+\tau^-,W^+W^-,$ and $ZZ$. }
   \label{fig:ratio}
 \end{center}
\end{figure}


Fig.~\ref{fig:ratio} shows the ratio of the shower/total muon flux
for various final states: $\nu \bar \nu, \mu^+\mu^-,\tau^+\tau^-,W^+W^-,$ and $ZZ$.
This ratio is not sensitive to the neutrino flavors: $\nu_e \bar \nu_e,\nu_\mu \bar \nu_\mu$ and $\nu_\tau \bar \nu_\tau$ yield a common ratio.
It is seen that the shower/total muon ratio is quite sensitive to the 
dark matter mass irrespective to the dark matter annihilation/decay modes.
Since this ratio expected from the atmospheric neutrino background is about 0.12,
the dark matter signals predict clearly large shower/total muon ratio.
Thus searching for muons signals with distinguishing showering and non-showering muons from 
particular direction may be useful for discriminating the signal from the background and
constraining/extracting properties of dark matter.

\section{Conclusions} \label{sec:conc}

In this paper, we have studied how dark-matter originated neutrinos can be detected
and how can we extract useful information on the dark matter properties,
motivated by recent observations of anomalous cosmic-ray positron/electron fluxes.

For this purpose,
we have focused on the separation between showering and non-showering upward muon fluxes
inside the neutrino detector.
Since muons with energy larger than about 500~GeV lose their energy dominantly by radiative processes,
such as bremmstrahlung and pair creation processes, 
searching for electromagnetic shower is a useful way to discriminate signals of high-energy neutrinos from atmospheric neutrino background.
Moreover, since the energy dependence of the total muon flux and showering muon flux
are different from each other, 
measuring the ratio between them gives us information on the dark matter mass.
A good point of muon search compared with gamma-ray and positron/electron flux is that
the muon flux is enhanced for heavier dark matter due to the enhancement of the
muon production and propagation distance inside the Earth.
Thus our method has a particular benefit for rather heavy dark matter case,
as is favored from the recent observations of the cosmic ray electron flux by the Fermi satellite.
The previous bound on the neutrino flux from the Galactic center given in Ref.~\cite{Desai:2004pq} comes from data of SK-I. Since the SuperKamiokande accumulates more data now, the bound is expected to be improved soon. 

Finally we mention possible applications to the on-going or future neutrino detectors.
The IceCube detector cannot see the upward muon signals from the Galactic center, since it is
located at the South pole.
A planned extension, called IceCube DeepCore~\cite{Cowen:2008zz}, may have a potential to 
search Galactic center by looking at downward muons
if atmospheric muons can be removed to the level of expected signals.
Although the sensitivity of the DeepCore on
the muon flux from the Galactic center is difficult to estimate at the
present stage, other targets such as dwarf galaxies may be used as a source of 
high-energy neutrinos.
Future megaton scale water tank detector, Hyper-Kamiokande, and
kilometer size detector, such as KM3NeT, is expected to significantly improve current
sensitivities of SK and may be ideal experiments to search for the dark matter signatures
by looking at the Galactic center.
Detailed estimates on the sensitivities in these experiments are beyond the scope of this paper,
and we leave this issue for future work.

\begin{acknowledgements}
J.H. appreciates Dr.~Kaneyuki and Dr.~Ibarra for useful discussion.
K.N. would like to thank the Japan Society for the Promotion of Science for financial support.
This work is supported by Grant-in-Aid
for Scientific research from the Ministry of Education, Science,
Sports, and Culture (MEXT), Japan, No.\ 20244037, No.\ 2054252 (J.H.),
and also by World Premier International Research Center Initiative, MEXT, Japan.

\end{acknowledgements}

{}


\begin{thebibliography}{}



\bibitem{Adriani:2008zr}
  O.~Adriani {\it et al.}  [PAMELA Collaboration],
  Nature {\bf 458}, 607 (2009)
  [arXiv:0810.4995 [astro-ph]].
   
  
\bibitem{:2008zz}
  J.~Chang {\it et al.},
  Nature {\bf 456}, 362 (2008).
  
  
\bibitem{Torii:2008xu}
  S.~Torii {\it et al.},
  arXiv:0809.0760 [astro-ph].
  
  
\bibitem{Collaboration:2009zk}
  Fermi~Collaboration,
  arXiv:0905.0025 [astro-ph.HE].
  
  
\bibitem{Aharonian:2009ah}
  H.~E.~S.~S. Collaboration, ~F.~Aharonian {\it et al.},
  arXiv:0905.0105 [astro-ph.HE].
  
  
\bibitem{Atoian:1995ux}
  A.~M.~Atoian, F.~A.~Aharonian and H.~J.~Volk,
  Phys.\ Rev.\  D {\bf 52}, 3265 (1995).
  
 
\bibitem{Hooper:2008kg}
  D.~Hooper, P.~Blasi and P.~D.~Serpico,
  JCAP {\bf 0901}, 025 (2009)
  [arXiv:0810.1527 [astro-ph]];
  H.~Yuksel, M.~D.~Kistler and T.~Stanev,
  arXiv:0810.2784 [astro-ph];
  K.~Ioka,
  arXiv:0812.4851 [astro-ph];
  N.~J.~Shaviv, E.~Nakar and T.~Piran,
  arXiv:0902.0376 [astro-ph.HE];
  Y.~Fujita, K.~Kohri, R.~Yamazaki and K.~Ioka,
  arXiv:0903.5298 [astro-ph.HE].
  
   
\bibitem{Bergstrom:2008gr}
  L.~Bergstrom, T.~Bringmann and J.~Edsjo,
  Phys.\ Rev.\  D {\bf 78}, 103520 (2008)
  [arXiv:0808.3725 [astro-ph]];
  M.~Cirelli and A.~Strumia,
  arXiv:0808.3867 [astro-ph];
  V.~Barger, W.~Y.~Keung, D.~Marfatia and G.~Shaughnessy,
  Phys.\ Lett.\  B {\bf 672}, 141 (2009)
  [arXiv:0809.0162 [hep-ph]];
  C.~R.~Chen, F.~Takahashi and T.~T.~Yanagida,
  Phys.\ Lett.\  B {\bf 671}, 71 (2009)
  [arXiv:0809.0792 [hep-ph]];
  I.~Cholis, L.~Goodenough, D.~Hooper, M.~Simet and N.~Weiner,
  arXiv:0809.1683 [hep-ph];
  M.~Cirelli, M.~Kadastik, M.~Raidal and A.~Strumia,
  Nucl.\ Phys.\  B {\bf 813}, 1 (2009)
  [arXiv:0809.2409 [hep-ph]].
  N.~Arkani-Hamed, D.~P.~Finkbeiner, T.~R.~Slatyer and N.~Weiner,
  Phys.\ Rev.\  D {\bf 79}, 015014 (2009)
  [arXiv:0810.0713 [hep-ph]];
  M.~Pospelov and A.~Ritz,
  Phys.\ Lett.\  B {\bf 671}, 391 (2009)
  [arXiv:0810.1502 [hep-ph]];
  C.~R.~Chen and F.~Takahashi,
  JCAP {\bf 0902}, 004 (2009)
  [arXiv:0810.4110 [hep-ph]];
  I.~Cholis, D.~P.~Finkbeiner, L.~Goodenough and N.~Weiner,
  arXiv:0810.5344 [astro-ph];
  Y.~Nomura and J.~Thaler,
  arXiv:0810.5397 [hep-ph];
  R.~Harnik and G.~D.~Kribs,
  arXiv:0810.5557 [hep-ph];
  D.~Feldman, Z.~Liu and P.~Nath,
  Phys.\ Rev.\  D {\bf 79}, 063509 (2009)
  [arXiv:0810.5762 [hep-ph]];
  P.~f.~Yin, Q.~Yuan, J.~Liu, J.~Zhang, X.~j.~Bi and S.~h.~Zhu,
  Phys.\ Rev.\  D {\bf 79}, 023512 (2009)
  [arXiv:0811.0176 [hep-ph]];
  K.~Ishiwata, S.~Matsumoto and T.~Moroi,
  arXiv:0811.0250 [hep-ph];
  Y.~Bai and Z.~Han,
  arXiv:0811.0387 [hep-ph];
  P.~J.~Fox and E.~Poppitz,
  arXiv:0811.0399 [hep-ph];
  K.~Hamaguchi, E.~Nakamura, S.~Shirai and T.~T.~Yanagida,
  arXiv:0811.0737 [hep-ph];
  A.~Ibarra and D.~Tran,
  JCAP {\bf 0902}, 021 (2009)
  [arXiv:0811.1555 [hep-ph]];
  I.~Cholis, G.~Dobler, D.~P.~Finkbeiner, L.~Goodenough and N.~Weiner,
  arXiv:0811.3641 [astro-ph].
  
  
\bibitem{Meade:2009iu}
  P.~Meade, M.~Papucci, A.~Strumia and T.~Volansky,
  arXiv:0905.0480 [hep-ph].
    

\bibitem{Kawasaki:1995cy}
  M.~Kawasaki, T.~Moroi and T.~Yanagida,
  Phys.\ Lett.\  B {\bf 370}, 52 (1996)
  [arXiv:hep-ph/9509399].


\bibitem{Moroi:1999zb}
  T.~Moroi and L.~Randall,
  Nucl.\ Phys.\  B {\bf 570}, 455 (2000)
  [arXiv:hep-ph/9906527];
  M.~Fujii and K.~Hamaguchi,
  Phys.\ Lett.\  B {\bf 525}, 143 (2002)
  [arXiv:hep-ph/0110072];
  Phys.\ Rev.\  D {\bf 66}, 083501 (2002)
  [arXiv:hep-ph/0205044];
  M.~Nagai and K.~Nakayama,
  Phys.\ Rev.\  D {\bf 76}, 123501 (2007)
  [arXiv:0709.3918 [hep-ph]];
  B.~S.~Acharya, P.~Kumar, K.~Bobkov, G.~Kane, J.~Shao and S.~Watson,
  JHEP {\bf 0806}, 064 (2008)
  [arXiv:0804.0863 [hep-ph]].
  
  
\bibitem{Profumo:2004ty}
  S.~Profumo and P.~Ullio,
  JCAP {\bf 0407}, 006 (2004)
  [arXiv:hep-ph/0406018];
  M.~Nagai and K.~Nakayama,
  Phys.\ Rev.\  D {\bf 78}, 063540 (2008)
  [arXiv:0807.1634 [hep-ph]];
  P.~Grajek, G.~Kane, D.~J.~Phalen, A.~Pierce and S.~Watson,
  arXiv:0807.1508 [hep-ph].


\bibitem{Hisano:2003ec}
  J.~Hisano, S.~Matsumoto and M.~M.~Nojiri,
  Phys.\ Rev.\ Lett.\  {\bf 92}, 031303 (2004)
  [arXiv:hep-ph/0307216];
  J.~Hisano, S.~Matsumoto, M.~M.~Nojiri and O.~Saito,
  Phys.\ Rev.\  D {\bf 71}, 063528 (2005)
  [arXiv:hep-ph/0412403].
  
  
\bibitem{Bertone:2008xr}
  G.~Bertone, M.~Cirelli, A.~Strumia and M.~Taoso,
  JCAP {\bf 0903}, 009 (2009)
  [arXiv:0811.3744 [astro-ph]];
  E.~Nardi, F.~Sannino and A.~Strumia,
  JCAP {\bf 0901}, 043 (2009)
  [arXiv:0811.4153 [hep-ph]];
  M.~Kawasaki, K.~Kohri and K.~Nakayama,
  arXiv:0904.3626 [astro-ph.CO];
  M.~Cirelli and P.~Panci,
  arXiv:0904.3830 [astro-ph.CO].
  
  
\bibitem{Donato:2008jk}
  F.~Donato, D.~Maurin, P.~Brun, T.~Delahaye and P.~Salati,
  Phys.\ Rev.\ Lett.\  {\bf 102}, 071301 (2009)
  [arXiv:0810.5292 [astro-ph]];
  P.~Grajek, G.~Kane, D.~Phalen, A.~Pierce and S.~Watson,
  arXiv:0812.4555 [hep-ph].
  
  
\bibitem{Braeuninger:2009pe}
  C.~B.~Braeuninger and M.~Cirelli,
  arXiv:0904.1165 [hep-ph];
  A.~Ibarra and D.~Tran,
  arXiv:0904.1410 [hep-ph].
  
  
\bibitem{Cumberbatch:2009ji}
  D.~T.~Cumberbatch, J.~Zuntz, H.~K.~K.~Eriksen and J.~Silk,
  arXiv:0902.0039 [astro-ph.GA];
  J.~Zhang, X.~J.~Bi, J.~Liu, S.~M.~Liu, P.~f.~Yin, Q.~Yuan and S.~H.~Zhu,
  arXiv:0812.0522 [astro-ph];
  K.~Ishiwata, S.~Matsumoto and T.~Moroi,
  Phys.\ Rev.\  D {\bf 79}, 043527 (2009)
  [arXiv:0811.4492 [astro-ph]];
  L.~Bergstrom, G.~Bertone, T.~Bringmann, J.~Edsjo and M.~Taoso,
  arXiv:0812.3895 [astro-ph].
  
  
\bibitem{Jedamzik:2004ip}
  K.~Jedamzik,
  Phys.\ Rev.\  D {\bf 70}, 083510 (2004)
  [arXiv:astro-ph/0405583].
  
   
\bibitem{Hisano:2008ti}
  J.~Hisano, M.~Kawasaki, K.~Kohri and K.~Nakayama,
  Phys.\ Rev.\  D {\bf 79}, 063514 (2009)
  [arXiv:0810.1892 [hep-ph]];
  J.~Hisano, M.~Kawasaki, K.~Kohri, T.~Moroi and K.~Nakayama,
  Phys.\ Rev.\ D {\bf 79}, 083522 (2009) [arXiv:0901.3582 [hep-ph]].
  
\bibitem{Galli:2009zc}
A.~V.~Belikov and D.~Hooper,
  arXiv:0904.1210 [hep-ph];
  S.~Galli, F.~Iocco, G.~Bertone and A.~Melchiorri,
  arXiv:0905.0003 [astro-ph.CO].
    
\bibitem{Hisano:2008ah}
  J.~Hisano, M.~Kawasaki, K.~Kohri and K.~Nakayama,
  Phys.\ Rev.\ D {\bf 79}, 043516 (2009) [arXiv:0812.0219 [hep-ph]].


\bibitem{Liu:2008ci}
  J.~Liu, P.~f.~Yin and S.~h.~Zhu,
  arXiv:0812.0964 [astro-ph].
  
  
\bibitem{Desai:2004pq}
  S.~Desai {\it et al.}  [Super-Kamiokande Collaboration],
  Phys.\ Rev.\  D {\bf 70}, 083523 (2004)
  [Erratum-ibid.\  D {\bf 70}, 109901 (2004)]
  [arXiv:hep-ex/0404025].

  
\bibitem{Desai:2007ra}
  S.~Desai {\it et al.}  [Super-Kamiokande Collaboration],
  Astropart.\ Phys.\  {\bf 29}, 42 (2008)
  [arXiv:0711.0053 [hep-ex]].
  
  
\bibitem{Ritz:1987mh}
  S.~Ritz and D.~Seckel,
  Nucl.\ Phys.\  B {\bf 304}, 877 (1988).
  
  
\bibitem{Kamionkowski:1991nj}
  M.~Kamionkowski,
  Phys.\ Rev.\  D {\bf 44}, 3021 (1991).
  
  
\bibitem{Mardon:2009rc}
  J.~Mardon, Y.~Nomura, D.~Stolarski and J.~Thaler,
  arXiv:0901.2926 [hep-ph].
    
  
\bibitem{Sjostrand:2006za}
  T.~Sjostrand, S.~Mrenna and P.~Skands,
  JHEP {\bf 0605}, 026 (2006)
  [arXiv:hep-ph/0603175].
  
  
\bibitem{Navarro:1995iw}
  J.~F.~Navarro, C.~S.~Frenk and S.~D.~M.~White,
  Astrophys.\ J.\  {\bf 462}, 563 (1996)
 [arXiv:astro-ph/9508025].


\bibitem{Burkert:1995yz}
  A.~Burkert,
  Astrophys.\ J.\  {\bf 447}, L25 (1995)
  [arXiv:astro-ph/9504041];
  P.~Salucci and A.~Burkert,
   Astrophys.\ J.\  {\bf 537}, L9 (2000).
   

\bibitem{Barger:2007xf}
  V.~Barger, W.~Y.~Keung, G.~Shaughnessy and A.~Tregre,
  Phys.\ Rev.\  D {\bf 76}, 095008 (2007)
  [arXiv:0708.1325 [hep-ph]].
  
  
\bibitem{Dutta:2000hh}
  S.~I.~Dutta, M.~H.~Reno, I.~Sarcevic and D.~Seckel,
  Phys.\ Rev.\  D {\bf 63}, 094020 (2001)
  [arXiv:hep-ph/0012350].
  
  
\bibitem{Honda:2004yz}
  M.~Honda, T.~Kajita, K.~Kasahara and S.~Midorikawa,
  Phys.\ Rev.\  D {\bf 70}, 043008 (2004)
  [arXiv:astro-ph/0404457].
  
  
\bibitem{Honda:2006qj}
  M.~Honda, T.~Kajita, K.~Kasahara, S.~Midorikawa and T.~Sanuki,
  Phys.\ Rev.\  D {\bf 75}, 043006 (2007)
  [arXiv:astro-ph/0611418].
  
  
\bibitem{Cowen:2008zz}
  D.~F.~Cowen  [IceCube Collaboration],
  J.\ Phys.\ Conf.\ Ser.\  {\bf 110}, 062005 (2008).

  

\end{thebibliography}
\end{document}